\documentclass{cernyrep} 
\usepackage{texnames}
\usepackage[T1]{fontenc}
\newcommand{\be}{\begin{equation}} \newcommand{\ee}{ \end{equation}}
\newcommand{\ba}{\begin{eqnarray}} \newcommand{\ea}{ \end{eqnarray}}

\begin{document}
\title{The Statistical Model of Nuclear Reactions: Open Problems}
 
\author{P. Fanto$^1$, Y. Alhassid$^1$, and H. A. Weidenm{\"u}ller$^2$}

\institute{$^1$Center for Theoretical Physics, Sloane Physics
  Laboratory, Yale University, New Haven, Connecticut 06520, USA \\
  $^2$Max-Planck-Institut f{\"u}r Kernphysik, D-69029 Heidelberg,
  Germany}

\maketitle % this produces the title block

\begin{abstract}
Several experiments~\cite{Koe10, Koe11, Koe13} show significant 
deviations from predictions of the statistical model of nuclear
reactions. We summarize unsuccessful recent theoretical efforts to
account for such disagreement in terms of a violation of orthogonal
invariance caused by the Thomas-Ehrman shift. We report on numerical
simulations involving a large number of gamma decay channels that also
give rise to violation of orthogonal invariance but likewise do not account for the
discrepancies. We discuss the statistical model in the light of these
results.
\end{abstract}

\section{Motivation}
\label{mot}

In recent years, several predictions of the statistical model of
nuclear reactions have been tested experimentally, with puzzling
results. The statistical model predicts that the reduced partial
neutron widths of isolated compound-nucleus resonances have a
Porter-Thomas distribution (PTD), i.e., a $\chi^2$ distribution in one
degree of freedom. Moreover, the total gamma decay width of an
isolated neutron resonance is the sum of a very large number of
partial gamma decay widths. If the latter have a PTD, the distribution
of total gamma decay widths should be very narrow.  However, the data
show strong deviations from these predictions:

\noindent (i) In the scattering of slow neutrons on the target nuclei
$^{192}$Pt and $^{194}$Pt, 158 and 411 isolated resonances,
respectively, were analysed. The data reject the validity of the PTD
with $99.997\%$ statistical significance~\cite{Koe10}.

\noindent (ii) A reanalysis of the nuclear data ensemble (NDE) rejects
the validity of the PTD with $99.97\%$ statistical
significance~\cite{Koe11}.

\begin{figure}
\centering
\includegraphics[width=0.5\textwidth]{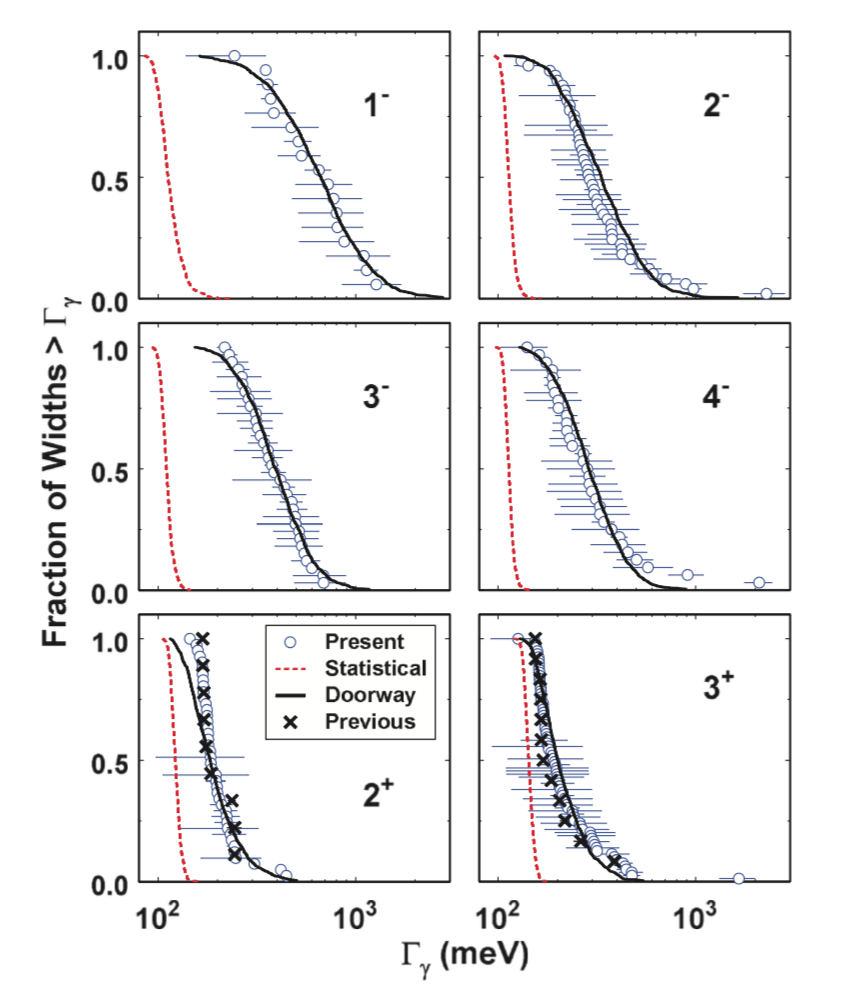}
\caption{\label{fig1} Cumulative distributions of the total gamma decay
  widths for neutron resonances in $^{96}$Mo as described in the
  text. Taken from Ref.~\cite{Koe13}.}
\end{figure}

\noindent (iii) The measured distributions of total gamma decay widths
for isolated neutron resonances in the compound nucleus $^{96}$Mo are
much wider and are peaked at significantly larger values of the widths
than predicted by the statistical model. The ground state of the
nucleus $^{95}$Mo has spin/parity $5/2^+$, hence $s$-wave and $p$-wave
resonances in the compound nucleus $^{96}$Mo have spin/parity values
$2^+, 3^+$ and $1^-, 2^-, 3^-, 4^-$, respectively. For all these
spin/parity values, Fig.~\ref{fig1} shows the measured cumulative
width distributions (i.e., the fraction of widths larger than a given
value) as dark lines with error bars and the values predicted by the
statistical model as red lines~\cite{Koe13}.

Here we summarize previous attempts and present a new approach aimed
at reconciling the statistical model with the data of
Refs.~\cite{Koe10, Koe11, Koe13}. To that end, we first recall in
Sec.~\ref{stat} the essential features of the statistical model.

\section{Statistical Model}
\label{stat}

For states of fixed spin and parity, we consider a compound-nucleus
(CN) reaction induced by slow neutrons ($s$-wave or $p$-wave) that
leads either to elastic neutron scattering or to gamma decay of the
CN. We neglect direct reactions. Denoting by $c = 1$ the neutron
channel and $c = 2, 3, \ldots, \Lambda \gg 1$ the gamma channels, the
scattering matrix is
\be
S_{c c'}(E) = \delta_{c c'} - 2 i \pi \sum_{\mu \nu} W_{c \mu}(E)
[( E - H^{\rm eff})^{- 1}]_{\mu \nu} W_{\nu c'}(E) \;.
\label{1}
\ee
Here $E$ is the excitation energy of the CN, with $E = 0$ for the
ground state. The threshold energy of channel $c$ is denoted by $E_c$.
The real matrix elements $W_{c \mu}(E) = W_{\mu c}(E)$ (defined for $E
\geq E_c$) describe the coupling of channel $c$ to the compound
nucleus states, spanned by the states $\mu = 1, 2, \ldots, N \gg
1$. In the framework of the statistical model, the nuclear Hamiltonian
is, within the space of compound states, replaced by $H^{\rm GOE}$, a
matrix of dimension $N \gg 1$ drawn from the Gaussian orthogonal
ensemble (GOE) of random matrices~\cite{Mit10}.  The effective
Hamiltonian is given by
\be
H^{\rm eff}_{\mu \nu} = H^{\rm GOE}_{\mu \nu} + \sum_c {\cal P}
\int_{E_c}^\infty {\rm d} E' \ \frac{W_{\mu c}(E') W_{c \nu}(E')}
{E -  E'} - i \pi \sum_c W_{\mu c}(E) W_{c \nu}(E) \;,
\label{2}
\ee
where $\cal P$ denotes the principal-value integral. 

The derivation of the PTD rests on the fact that the GOE is invariant
under orthogonal transformations in the space of compound states. Such
invariance is violated by the last two terms on the right-hand side of
Eq.~(\ref{2}), describing the coupling of the CN to the channels. In
recent years, considerable theoretical effort has been devoted to the
question whether the discrepancies listed in Sec.~\ref{mot} can be
attributed to these two terms.

We discuss these terms using simplifications that apply in the present
case. (i) For the overwhelming majority of gamma channels ($c \geq 2$)
the gamma decay energy $E$ differs substantially from the threshold
energy $E_c$. Then it is legitimate to neglect the principal-value
integrals, and we do so for all gamma channels. (ii) For slow neutrons
the energy dependence of the matrix elements that couple to the
neutron channel is given by $W_{1 \mu}(E) \approx (E - E_1)^{(2 l +
  1)/4} W^{(0)}_{1 \mu}$. Here $l = 0$ ($l = 1$) for $s$-wave
($p$-wave) neutrons, respectively, while $W^{(0)}_{1 \mu}$ is
independent of energy. With that approximation the two coupling terms
in Eq.~(\ref{2}) are given by the product of $W^{(0)}_{\mu 1}
W^{(0)}_{1 \nu}$ and of a complex energy-dependent factor that is
independent of $\mu, \nu$.  (iii) The absence of direct reactions
implies channel orthogonality in the form $\pi \sum_\mu W_{c \mu}
W_{\mu c'} = \delta_{c c'} \lambda \kappa_c$. Here $\lambda = N d /
\pi$ is the standard GOE parameter and $d$ is the mean GOE level
spacing in the center of the spectrum. The dimensionless parameter
$\kappa_c$ measures the strength of the coupling to channel
$c$. Channel orthogonality allows us to diagonalize the matrices
coupling the compound states to the channels, and we obtain
\be
H^{\rm eff}_{\mu \nu} = H^{\rm GOE}_{\mu \nu} + \delta_{\mu \nu} V_\mu \ ,
\label{3}
\ee
where
\ba
V_c = \left\{ \begin{array}{cl} \lambda \bigg( \frac{1}{\pi} {\cal P} \int_{E_1}^\infty {\rm
  d} E' \frac{\kappa_1(E')}{E - E'} - i \kappa_1(E) \bigg) & {\rm for} \ c = 1\;, \\
 - i \lambda \kappa_c & {\rm for} \ c = 2, \ldots, \Lambda \;, \\
 0 \ & {\rm for} \ c > \Lambda \;.
 \end{array}
 \right.
\label{4}
\ea
The average $S$ matrix is $\langle S_{c c'} \rangle = \delta_{c c'} (1
- \kappa_c)/(1 + \kappa_c)$, and the transmission coefficient in
channel $c$ is $T_c = 1 -|\langle S_{cc}\rangle|^2 = 4 \kappa_c / (1 +
\kappa_c)^2$.

\section{Nonstatistical Effects: Thomas-Ehrman Shift}

In the platinum isotopes, the single-particle $4 s$ state of the
nuclear shell model is close to neutron threshold, causing a maximum
of the $s$-wave neutron strength function in that mass region and, at
the same time, an enhancement of the principal-value integral (the
shift function) in $V_1$ of Eq.~(\ref{4}). In light nuclei the shift
due to the principal-value integral is known as the Thomas-Ehrman
shift.  Several authors~\cite{Vol15, Bog17, Fan17} have addressed the
question whether that enhancement may be responsible for deviations of
the distribution of partial neutron resonance widths from the PTD. In
all these works, the gamma channels were neglected (except for a
constant imaginary shift in the GOE energies).

For real values of $V_1$ that are consistent with the enhancement due
to the $4 s$ state but were taken to be independent of energy, the
numerical results of Volya et al.~\cite{Vol15} showed significant
deviations of the distribution of reduced partial neutron widths from
the PTD. Bogomolny~\cite{Bog17} succeeded in diagonalizing the
Hamiltonian~(\ref{3}) with an energy-independent real $V_1$ in the
limit of large matrix dimension $N$. He showed that deviations from
the PTD do arise but that locally the width distribution remains a
PTD. The effect found in Ref.~\cite{Vol15} is attributed to ignoring
the secular variation of the average width with energy. Once the
average width is divided out, the fluctuations of the reduced widths
follow the PTD. Fanto et. al.~\cite{Fan17} studied a model that
includes a realistic description of the neutron channel in a
Woods-Saxon potential, and effectively takes full account of the
energy dependence of $V_1$. The resulting local distribution of
partial neutron widths was found to be consistent with the PTD.

We note that a sufficiently large imaginary part of $V_1$ can lead to
deviations from a PTD~\cite{Cel2011}. However, no such deviations are
observed for realistic values of $\kappa_1$~\cite{Vol15,Fan17}.

Therefore, the deviations from the PTD listed under (i) and (ii) in
Sec.~\ref{mot} cannot, for realistic values of the parameters, be
accounted for by violations of orthogonal invariance due to the
coupling to the neutron channel. In view of this result, we consider
the shift function in the first of Eqs.~(\ref{4}) to be insignificant
and we disregard it in what follows.

\section{Nonstatistical Effects: Many Gamma Channels}

In medium-mass and heavy nuclei, the number of gamma decay channels is
huge (of the order of $10^6$ or so) for each isolated neutron
resonance. We ask: can that fact account for the deviations listed in
Sec.~\ref{mot} even though the coupling of each individual gamma
channel to the CN is very weak? Prior to addressing that question we
recall in Sec.~\ref{sm} the calculation of the total gamma decay width
in the statistical model.

\subsection{Total Gamma Width}
\label{sm}

An isolated neutron resonance labelled $\mu$ with spin/parity $J^\pi$
and resonance energy $E_\mu$ decays by emission of photons of
multipolarity $L$ and parity $\pi$ (E1, M1, E2, M2, \ldots, jointly
written as $XL$) to final states $f$ with spins/parities
$J^\pi_f$. (The label $f$ replaces the channel label $c$ used in
Eq.~(\ref{3}).) The corresponding partial decay widths are denoted by
$\Gamma^{J^\pi}_{\mu \gamma J^\pi_f f XL}$. The total gamma decay
width is
\be
\Gamma^{J^\pi}_{\mu \gamma} = \sum_{XL} \sum_{J^\pi_f f}
\Gamma^{J^\pi}_{\mu \gamma J^\pi_f f XL}  \ .
\label{5}
\ee
In the statistical model, the partial width is written as a product,
\be
\Gamma^{J^\pi}_{\mu \gamma I^\pi_f f XL} = x^2_f
\langle \Gamma^{J^\pi}_{\mu \gamma J^\pi_f f XL} \rangle \ .
\label{6}
\ee
The average value in Eq.~(\ref{6}) is expressed in terms of the gamma
strength function $f_{XL}(E_\gamma)$ and the average spacing
$d_{J^\pi}$ of the resonances of spin $J$ and parity $\pi$,
\be
\langle \Gamma^{J^\pi}_{\mu \gamma I^\pi_f f XL} \rangle = d_{J^\pi}
E^{2 L + 1}_\gamma f_{XL}(E_\gamma) = d_{J^\pi} \frac{2}{\pi}
\kappa_f \;,
\label{7}
\ee
where the dimensionless parameters $\kappa_f$ have the same physical
meaning as the parameters $\kappa_c$ in Eq.~(\ref{4}). The factors
$x^2_f$ are uncorrelated random variables that each have mean value
unity and follow the PTD. These account for fluctuations of the
partial widths.  The statistical model is implemented by choosing
values for the strength function and for the average level density
from which the actual values of the final energies $E_f$ are drawn.
The average total total gamma decay width is given by 
\be
\langle \Gamma^{J^\pi}_{\mu \gamma} \rangle = d_{J^\pi} \sum_{XL} \sum_{J^\pi_f} \int_0^{E_\mu}
{\rm d} E_\gamma \ \rho(E_\mu - E_\gamma, J^\pi_f) E^{2 L + 1}_\gamma f_{XL}(E_\gamma) \;,
\label{8}
\ee
where $\rho(E_\mu - E_\gamma, J^\pi_f)$ is the average level density
at energy $E_\mu-E_\gamma$ and spin/parity $J^\pi_f$ of the final
levels into which the compound nucleus decays.

\subsection{Simulation of Gamma Decay of the $^{96}$Mo
Compound Nucleus}
\label{CN}

The influence of the coupling of many gamma channels on the
statistical properties of the neutron resonances was simulated as
follows. It is impractical to use in Eqs.~(\ref{3}) and (\ref{4}) the
totality of channels $f$ resulting from the treatment in
Section~\ref{sm}. Their number is simply too large. For each group of
neutron resonances carrying spin-parity values $(2^+, 3^+)$ for
$s$-wave neutrons and $(1^-, 2^-, 3^-, 4^-)$ for $p$-wave neutrons, a
set of representative channels $c = 1, 2, \ldots, \Lambda - 1$
(distinguished in notation from the actual channels labelled $f$) was
constructed instead. Below an excitation energy of $2.79$ MeV, known
measured discrete levels were used. Above that energy, representative
channels labelled $c$ were defined by coarse-graining: final states
$f$ close in energy and carrying identical quantum numbers were
grouped together. The average density of final states $c$ was taken
proportional to the actual level density. For the latter, the
back-shifted Fermi gas model with a spin distribution described by the
spin cutoff model were used. States with opposite parity were assumed
to have the same level density. Only E1 and M1 gamma transitions were
considered as these contribute the bulk to the widths. For the E1 and
M1 strength functions the parametrization of Ref.~\cite{She09} was
used. The effective coupling parameters $\kappa_c$ are sums over the
parameters $\kappa_f$ for the states $f$ in the group. The number of
gamma channels so constructed was $400$. The total number of channels
was 401: one neutron channel, 200 E1 channels, and 200 M1
channels. Results shown below are taken from the middle of the GOE
spectrum to avoid edge effects.

\subsection{Results}
\label{res}

\begin{figure}[h!]
\centering
\includegraphics[width=0.6\textwidth]{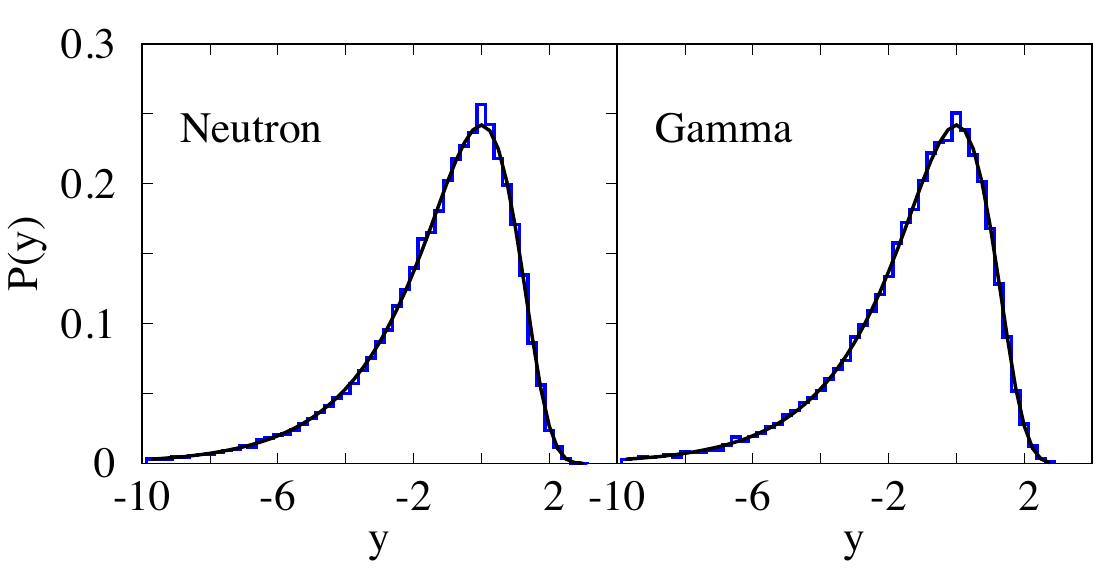}
\caption{\label{fig2} Distribution of $y = \ln x$, where $x =
  \Gamma/\langle \Gamma \rangle$ is the reduced partial
  width. The left panel shows the result for the neutron channel,
  the right panel shows the result for the most strongly
  coupled gamma channel. Histograms are model calculations (see text), and the
  solid black lines are the PTD.}
\end{figure}

The scheme described in Sec.~\ref{CN} was used to check for deviations
of neutron and gamma decay widths from the PTD. For a width $\Gamma$
and its average $\langle \Gamma \rangle$ we define $x = \Gamma /
\langle \Gamma \rangle$ and $y = \ln x$. For $y$ the PTD takes the
form $P(y) = \sqrt{x / (2 \pi)} \exp (- x / 2)$.  For neutron
resonances with spin/parity $2^+$ that function is shown in
Fig.~\ref{fig2} for the partial widths in the neutron channel and in
the most strongly coupled E1 channel. The solid black lines describe
the PTD, and the histograms are the result of the simulations. They
agree perfectly. We have found similar agreement for less strongly
coupled gamma channels and for other spin/parity values. We conclude
that a large number of gamma channels with realistic coupling
strengths does not alter the PTD of partial widths in any channel
(neutron or gamma).

\begin{figure}[bth]
\centering
\includegraphics[width=0.5\textwidth]{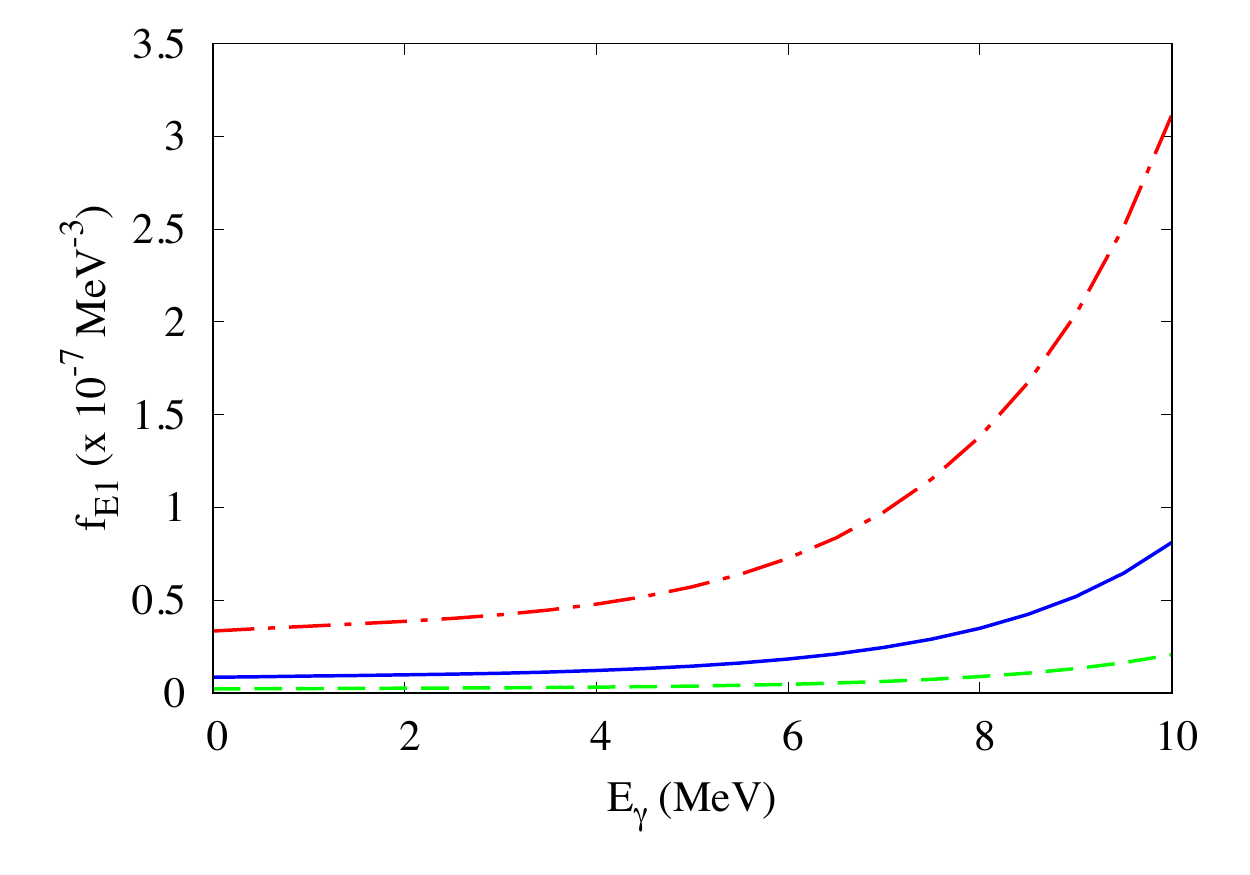}
\caption{\label{fig3} Three different parametrizations of the E1
  strength function as described in the text.}
\end{figure}

\begin{figure}[bth]
\centering
\includegraphics[width=0.5\textwidth]{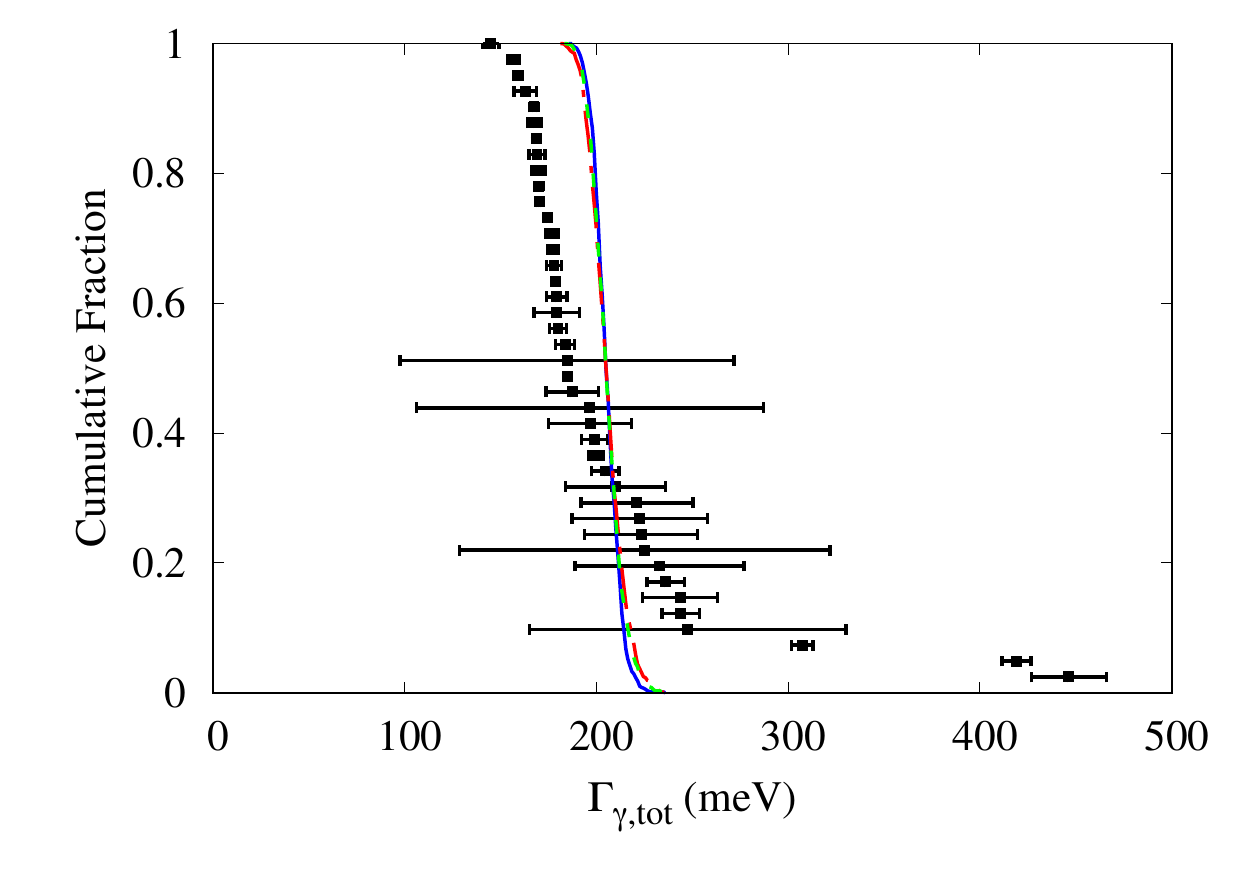}
\caption{\label{fig4} The cumulative distribution of the total gamma decay
  width for the $2^+$ resonances in $^{96}$Mo. The black squares with error bars are data taken
  from Ref.~\cite{Koe13}. The three colored lines correspond to simulations that use 
  the three parametrizations of the E1 strength function shown in Fig.~\ref{fig3}.}
\end{figure}

Total gamma decay widths were obtained from Eqs.~(\ref{5}), (\ref{6})
and (\ref{7}) with the adopted values for the level density and the
strength function, and with a PTD for the variables $x^2_f$. To check
for the sensitivity of our results to the form of the strength
function we have used three different parametrizations of the E1
strength function displayed in Fig.~\ref{fig3}.  The blue solid line describes
 the standard parameterization of the E1 strength function, the red dotted-dashed line 
 is found by taking twice the width of the giant dipole resonance in the E1
 strength function, and the green dashed line corresponds to half the width 
 of the giant dipole resonance. The resulting
cumulative distributions of the total gamma decay width for the $2^+$
neutron resonances (normalized to give the experimental total average
gamma width) are shown in Fig.~\ref{fig4}. 
%The blue solid line is obtained using the standard parameterization of the E1 strength
%function, the red dashed line is found by taking twice the width of the giant dipole
% resonance in the E1 strength function, and the green dotted-dashed corresponds 
%to  half the width (shown in Fig.~\ref{fig3}). 
These cumulative distributions virtually coincide. Also shown with error bars is the
measured cumulative distribution of Ref.~\cite{Koe13}. That
distribution is clearly much wider than the simulated ones.

The disagreement of statistical-model predictions with the data of
Ref.~\cite{Koe13} extends to the maxima of the (un-normalized) total
gamma decay width distributions.  Table~\ref{T1} shows that the
locations of the peaks of the distributions (or more accurately, the
average values) also differ markedly.  The experimental average total
widths have significantly larger values than the corresponding average
withs predicted by the statistical model. For the $1^-$ states the
ratio is larger than three. Individual peak positions for given
spin/parity can be fitted by an ad-hoc modification of the E1 strength
function. It was not possible, however, to fit all peak positions
simultaneously by such modification.

\begin{table}[h!]
\centering
\begin{tabular}{|c|c|c|c|c|c|c|}
\hline
$J^{\pi}$ & $2^+$ & $3^+$ & $1^-$ & $2^-$ & $3^-$ & $4^-$ \\ \hline
$\langle \Gamma_{\gamma,\rm sim} \rangle$ (meV) & 165.5 & 157.5 & 191.2 & 172.8 & 169.2 & 153.8 \\ \hline
$\langle \Gamma_{\gamma,\rm exp} \rangle$ (meV) & 206 (31) & 240 (58) & 670 (225) & 374 (115)  & 404 (100) & 361 (106)  \\ \hline
\end{tabular}
\caption{\label{T1} Comparison of simulated average total gamma widths
  $\langle \Gamma_{\gamma,\rm sim} \rangle$ with the experimental results
  $\langle \Gamma_{\gamma,\rm exp} \rangle$.}
\end{table}

\begin{figure}[h!]
\centering
\includegraphics[width=0.45\textwidth]{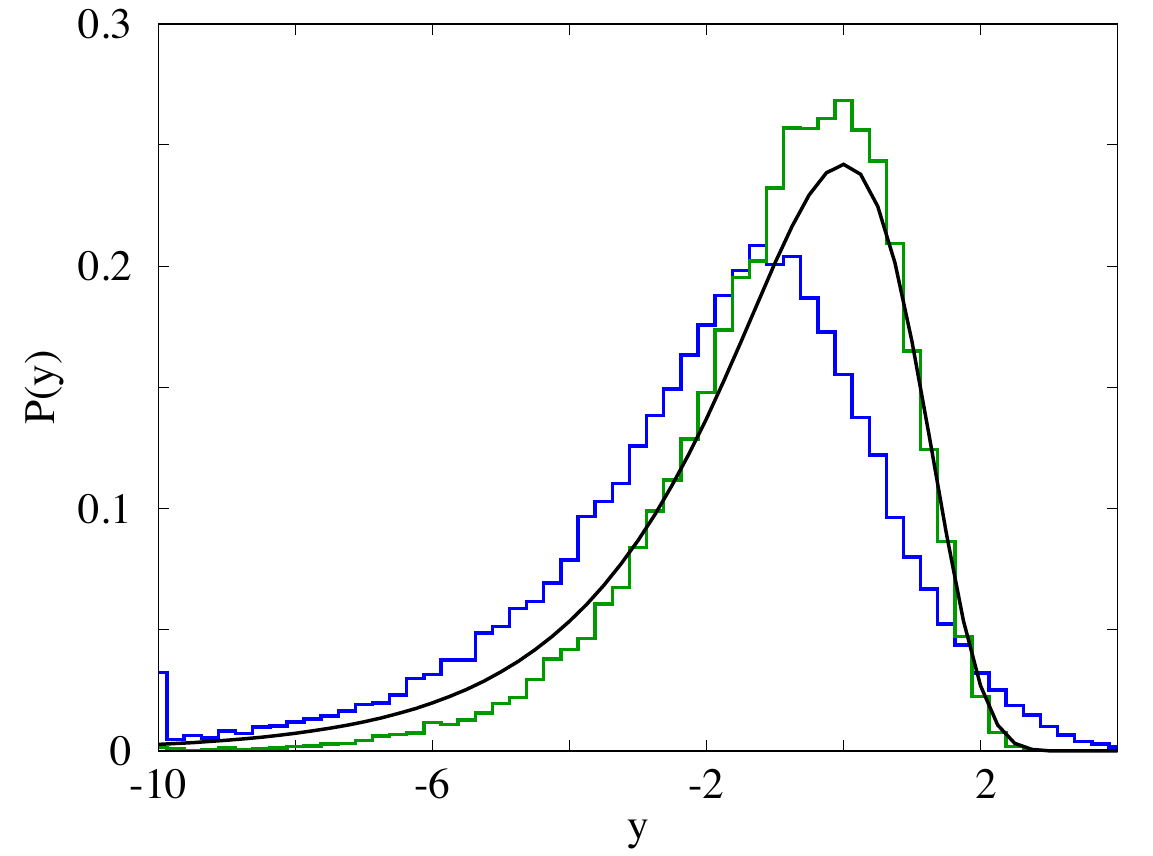}
\caption{\label{fig5} Distribution of $y_i = \ln x_i$ for the model of
  Eq.~(\ref{3}) with $V_\mu = - i 0.8 \lambda \delta_{\mu 1}$, where
  $i = 1$ denotes the neutron channel (blue histogram) and $i = 2$ denotes one gamma channel (green histogram).
  The set $\{x_i\}$ equals the set $\{\Gamma_{i \mu}/\langle
  \Gamma_{i\mu} \rangle)\}$, and $y_i = \ln x_i$ as in Fig.~\ref{fig2}.
  % The blue histogram is for the neutron  channel, and the green histogram is for one gamma channel. 
  The solid  black line is the PTD.}
\end{figure}

\begin{figure}[h!]
\centering
\includegraphics[width=0.5\textwidth]{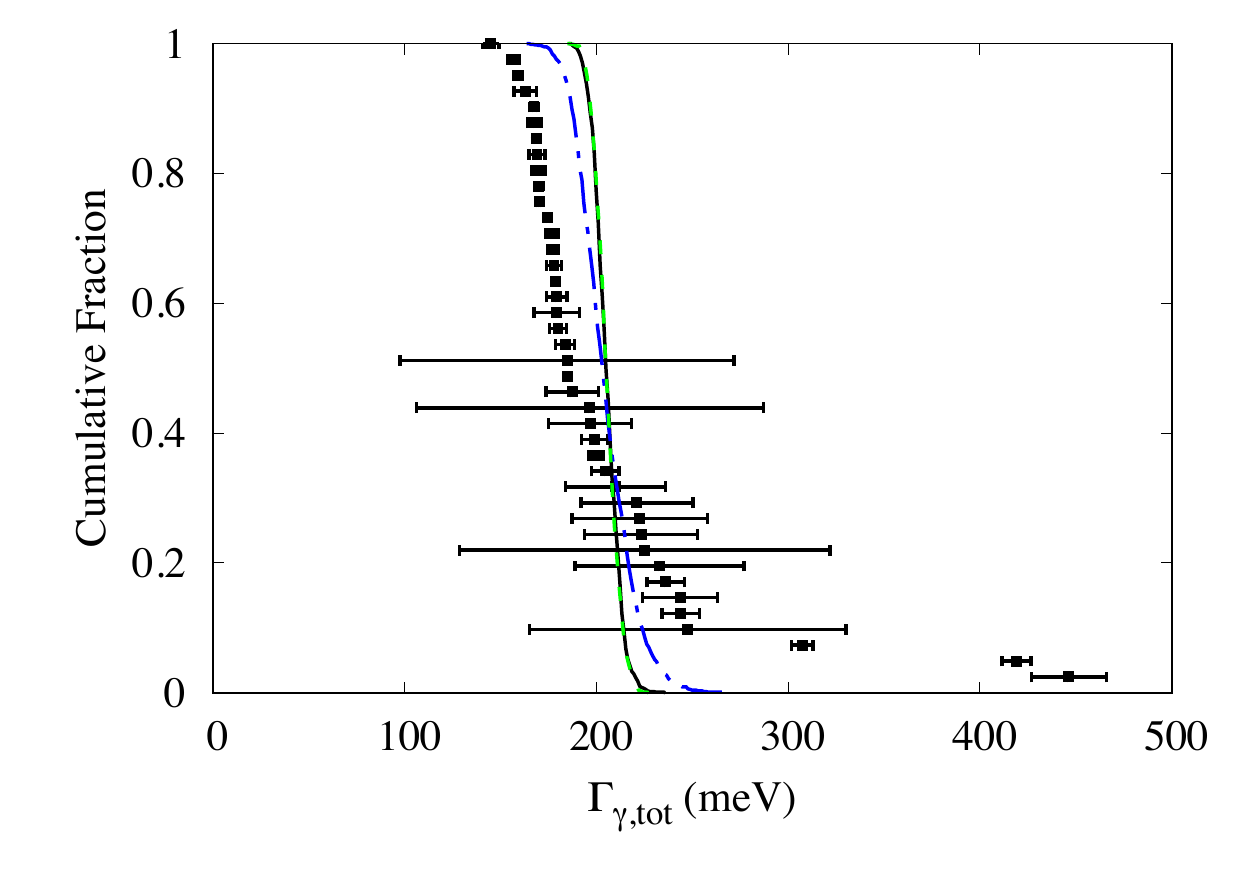}
\caption{\label{fig6} Simulated cumulative total width distributions
  are compared with experimental data for resonances of spin-parity
  $2^+$. The black solid line is obtained using the PTD for the
  partial widths. The green dashed line and blue dotted-dashed line
  are obtained using modified distributions for the partial gamma
  widths given, respectively, by the green and blue histograms in
  Fig.~\ref{fig5}. The black squares with error bars show the
  experimental data of Ref.~\cite{Koe13}. The simulated widths are
  normalized to match the experimental average total width.}
\end{figure}

The PTD for the partial gamma decay widths is corroborated by the
results shown in Fig.~\ref{fig2}. The disagreement of the simulated
neutron partial width distribution (a PTD) with the data of
Refs.~\cite{Koe10, Koe11} may cast doubt on the validity of that
conclusion. That raises the question: Would a distribution of partial
gamma widths different from the PTD yield better agreement of the
predicted distribution of total gamma widths with the data of
Ref.~\cite{Koe13}? To generate such a distribution we have used a
dynamical model with an unrealistically strong coupling of the GOE
Hamiltonian to the neutron channel, $V_\mu = - i 0.8 \lambda
\delta_{\mu 1}$ in Eq.~(\ref{3}).  The resulting distributions of the
partial widths are shown in Fig.~\ref{fig5} for the neutron channel
(blue histograms) and for one gamma channel (green histograms). Both
distributions differ markedly from the PTD (black). It is noteworthy
that strong coupling to the neutron channel also modifies the width
distribution in the gamma channels. However, Fig.~\ref{fig6} shows
that use of the modified distribution for the gamma channels does not
affect the disagreement with the data of Ref.~\cite{Koe13}.  The
calculated cumulative distribution for the $2^+$ resonances using the
modified gamma channel distribution (dashed green line) is nearly
indistinguishable from the one obtained using the PTD (black line).
Even if we use for the partial gamma width distribution the blue
distribution in Fig.~\ref{fig5} (i.e., assuming the gamma channels to
be strongly coupled to the resonances), the resulting cumulative
distribution (dashed-dotted blue line) is not much different. In
comparison, the data for the $2^+$ resonances (black squares with
error bars) differ substantially from the theoretical curves.

The reason for the near coincidence of the simulated cumulative
distributions is actually quite simple. According to the construction
in Eqs.~(\ref{5}) to (\ref{6}), the total gamma width is the sum of a
very large number ($K$, say) of independently distributed random
variables with similar or identical distributions (describing the
partial gamma widths). The central limit theorem implies that the
total width has a Gaussian distribution with a variance that is
inversely proportional to $K$. The narrowness of the predicted total
width distributions is, thus, independent of the actual form of the
partial-width distribution and is a universal feature resulting
directly from the basic tenets of the statistical model.

\section{Discussion}

The distributions of reduced partial neutron widths reported in
Refs.~\cite{Koe10, Koe11} deviate significantly from the PTD. Within
the framework of the statistical model, violation of orthogonal
invariance is a possible culprit. Two mechanisms for such violation
have been investigated. The Thomas-Ehrman shift, addressed by several
authors, is ruled out. Simulations involving a large number of gamma
channels yield perfect agreement with the PTD in all channels. That
mechanism is, therefore, also ruled out. Furthermore, the measured
distributions of total gamma decay widths in $^{96}$Mo reported in
Ref.~\cite{Koe13} disagree with those obtained in simulations that
follow the statistical model. These predicted distributions are too
narrow, and the mean values of the total gamma widths are too small
when compared with the experimental results.

In summary, the statistical model fails to account for the data in
platinum isotopes, in the nuclear data ensemble, and in
$^{96}$Mo. Violation of orthogonal invariance due to channel coupling
cannot be held responsible for these failures. The observed deviations
suggest that at neutron threshold, the mixing of CN states is less
complete than implied in the statistical model by the use of a GOE
Hamiltonian. The small mean values of the total gamma decay widths
predicted by the statistical model show that gamma strength is missing
in the model. That poses the question whether the Brink-Axel
hypothesis actually applies to all gamma transitions that contribute
significantly to the gamma decay of the neutron resonances. Such
consequences are rather drastic. We feel that additional experimental
tests of the statistical model are called for.

Given the experimental results, a small number of strong gamma
transitions to low-lying final states might offer a way out of the
dilemma~\cite{Koe13}. These should have sufficient intensity to shift
the simulated distribution of total gamma decay widths towards the
experimental values. Their number should be sufficiently small to
overcome the limits of the central-limit theorem and to broaden the
distribution of total gamma decay widths. It is an open question
whether a comparatively small number of such strongly coupled gamma
decay channels could also cause a change of the PTD for the reduced
neutron widths.

\section*{Acknowledgements}
We thank M. Krti\v{c}ka and P.~E. Koehler for useful discussions.  We
also thank P.~E. Koehler for providing the experimental data used
here.  This work was supported in part by the U.S.~DOE Grant
No.~DE-FG02-91ER40608 and by the U.S.~DOE NNSA Stewardship Science
Graduate Fellowship under cooperative agreement No.~DE-NA0003864.  The
initial part of this work was performed at the Aspen Center for
Physics, which is supported by a National Science Foundation Grant
No.~PHY-1607611.

\end{document}